\documentclass[reprint, aps,pra,notitlepage, nofootinbib,longbibliography,floatfix]{revtex4-2}

\usepackage[utf8]{inputenc}
\usepackage{float}
\usepackage{graphicx}  
\usepackage{physics}
\usepackage{esint}
\usepackage{amssymb}  
\usepackage{mathtools}
\usepackage{amsmath}
\usepackage{amsthm}
\usepackage{mathrsfs}
\usepackage[colorlinks=true,linkcolor=blue,citecolor=magenta,urlcolor=magenta,plainpages=false,pdfpagelabels]{hyperref}
\usepackage{color,xcolor,colortbl}
\usepackage{bm}
\usepackage[most]{tcolorbox}
\usepackage[normalem]{ulem}

\newcommand{\e}{\text{e}}
\newcommand{\im}{\text{i}}
\newcommand{\bea}{\begin{eqnarray}}
\newcommand{\eea}{\end{eqnarray}}

\begin{document}

\title{Stimulating the Quantum Aspects of an Optical Analog White-Black Hole}

\author{Ivan Agullo}
\email{agullo@lsu.edu}
\author{Anthony J. Brady}
\email{abrady6@lsu.edu}
\author{Dimitrios Kranas}
\email{dkrana1@lsu.edu}
\affiliation{Department of Physics and Astronomy, Louisiana State University, Baton Rouge, LA 70803, U.S.A.
}

%\date{\today}

\begin{abstract}
This work introduces a  synergistic combination of  analytical methods
and numerical simulations  to study the propagation of weak wave-packet modes in an optical medium containing the analog of a pair white-black hole. We apply  our tools to analyze several aspects of the evolution, such as (i) the region of the parameter space where the analogy with the Hawking effect is on firm ground and (ii) the influence that ambient thermal noise and detector inefficiencies have on the observability of the Hawking effect. We find that aspects of the Hawking effect that are of quantum origin, such as quantum entanglement, are extremely fragile to the influence of inefficiencies and noise.  
We propose a protocol to amplify and  observe these quantum aspects, based on seeding the process with a single-mode squeezed input.

\end{abstract}

\maketitle

{\bf Introduction.} 
The Hawking effect of spontaneous particle pair creation by black holes \cite{Hawking:1974rv,Hawking:1975vcx} can be understood as a process of two-mode quantum squeezing triggered by a causal horizon. What makes the phenomenon remarkable is not only the squeezing---which generically appears in other time-dependent spacetimes--- but its intrinsic thermal character, allowing one to associate a temperature with the horizon.
The connection with thermodynamics \cite{Bardeen:1973gs,Bekenstein:1974ax} and the associated universality of the Hawking effect \cite{Visser2003} lead to a profound and fertile crossroad between diverse areas of physics.
There is, therefore, a strong motivation to experimentally confirm this prediction, as well as to explore  open issues in Hawking's derivation, such as the role of arbitrarily high energy modes \cite{Jacobson:1991gr} or a potential loss of unitarity \cite{Hawking:1976ra}. This interest has motivated a plethora of analog models, in which the physics of squeezing generated by causal barriers can be recreated in the laboratory \cite{philbin08, weinfurtner2011,euve2016, steinhauer2016,de19BEC,drori19,kolobov2021BEC}.

A major challenge for observing aspects of the spontaneous Hawking process, even in analog models, is the extraordinarily weak character of the output, easily masked by ambient noise. At present, only one group has claimed success \cite{de19BEC, kolobov2021BEC}, and the community awaits for further confirmation. A promising alternative 
is to enhance the intensity of the output by replacing the initial vacuum state with a non-vacuum input, i.e., to focus on the {\em stimulated} Hawking process. Although stimulated Hawking radiation has been accessed in laboratory experiments \cite{weinfurtner2011,euve2016,drori19}, one can explain the observations made so far as a process of  classical amplification of waves.  Consequently, the stimulated process has been regarded as containing little value to assert the quantum nature of the Hawking effect \cite{weinfurtner2011,drori19}.  

The goal of this work is to introduce a  strategy to enhance the {\em quantum} aspects of the Hawking process. We are motivated by  the observation that stimulating the Hawking effect can also amplify the entanglement between the outputs---not merely their intensities---as long as one chooses appropriate quantum initial states and systematic inefficiencies are sufficiently under control. We describe a protocol to observe the amplified entanglement  and to unambiguously identify the main characteristics of the Hawking process and its quantum origin out of observations. 

Although the core of our ideas is general, we formulate them in the context of optical  systems, where the analog gravitational configuration is made of a pair white-black hole. The advantage is the possibilities optical systems  offer to generate, manipulate, and observe quantum states as well as their entanglement structure \cite{raymer2009}.
We use units in which $c=\hbar=k_B=1$.

{\bf Set up.} Optical systems provide a popular scenario to recreate the physics of the Hawking process \cite{philbin08, drori19,demircan11TRANSISTOR,rubino2012soliton,petev2013blackbody, rosenberg2020optical}. An electromagnetic pulse propagating in a dielectric  medium can locally change the  optical properties of the medium, modifying the refractive index (Kerr effect). In this way, by introducing strong pulses in non-linear materials, one can modify the speed of propagation of weak probes propagating  thereon. Probes that are initially faster than the  pulse will slow down when trying to overtake it, and if the pulse is strong enough, its rear end will act as an impenetrable (moving) barrier. This is the optical analog of the horizon of a white hole---a region where no signal can enter. Similarly, an analog black hole horizon appears in the front end of the pulse. Since the pair white-black hole propagates with the strong pulse, from now on we will work in the comoving frame.

\begin{figure}[t]
    {\centering     
\includegraphics[width = 0.48\textwidth]{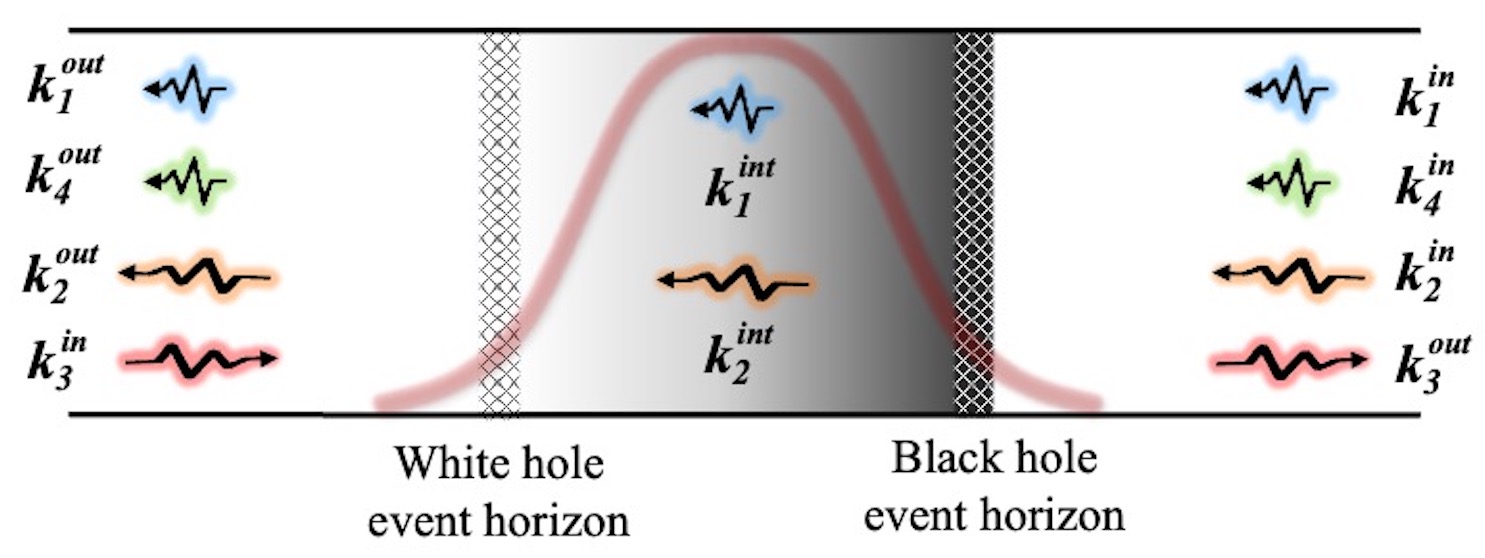}
}
\caption{Illustration of the structure of  $in$, $int$ and $out$ modes for an optical analog white-black hole in a comoving frame.}\label{modestructure}
\end{figure}

The presence of white-black horizons can also be understood by looking at the dispersion relation for  weak probes of frequency $\omega$. A detailed analysis of the dispersion relation of dielectric materials with a sub-luminal dispersion relation and characterized effectively by a single Sellmeier term, such as diamond, can be found in \cite{linder16}. The most relevant features are the following. 
Far away from the strong pulse, the dispersion relation has four solutions, $k_i$, $i=1 \dots 4$. The modes $k_1$ and $k_4$ are short-wavelength modes, in contrast to $k_2$ and $k_3$. Furthermore, $k_1,k_2$ and $k_4$ are left-movers (negative group velocity), while $k_3$ wave-packets propagate to the right (see Fig.~\ref{modestructure}). The strong pulse modifies the dispersion relation in such a way that, inside the pulse, the wavenumbers $k_3$ and $k_4$ become complex and no longer describe propagating modes. Only $k_1$ and $k_2$ propagate in the interior region, and since both are left-movers, they will necessarily exit the rear end of the pulse. Hence, the interior of the strong pulse is analogous to the interior region of a white-black hole pair. 

Since the properties of the medium are time-independent in the comoving frame, the frequency $\omega$ is conserved. The wavenumbers $k_i$, in contrast, are not conserved, due to the spatial dependence of the optical properties of the medium caused by the strong pulse. 
Hence, propagation across the white-black hole will generically mix the four modes  $k_i$. 
It is not difficult to see that the mode $k_1$ has a negative symplectic norm, in contrast to $k_2,k_3,k_4$ \cite{linder16}. As a consequence, the mixing of the four modes %caused by the evolution 
will cause a certain degree of squeezing and particle creation.

{\bf The analog quantum circuit.} Although the existence of optical horizons originates in non-linear optics, it is well known that the evolution of weak probes is well approximated by linear equations, and the non-linearities induced by the strong pulse  can be all encoded in the optical properties of the medium.  
This is the analog of the quantum field theory in curved spacetimes used in Hawking's original  derivation. In the optical set up, the resolution of the evolution reduces to computing the scattering matrix ($S$-matrix) describing the dynamics of wave-packet modes which, in the asymptotic region, have  wavenumber centered around $k^{in}_i$. Since different frequencies $\omega$ do not mix with each other, one computes the $S$-matrix for each individual frequency. 

\begin{figure}[t]
{\centering     
\includegraphics[width = 0.48\textwidth]{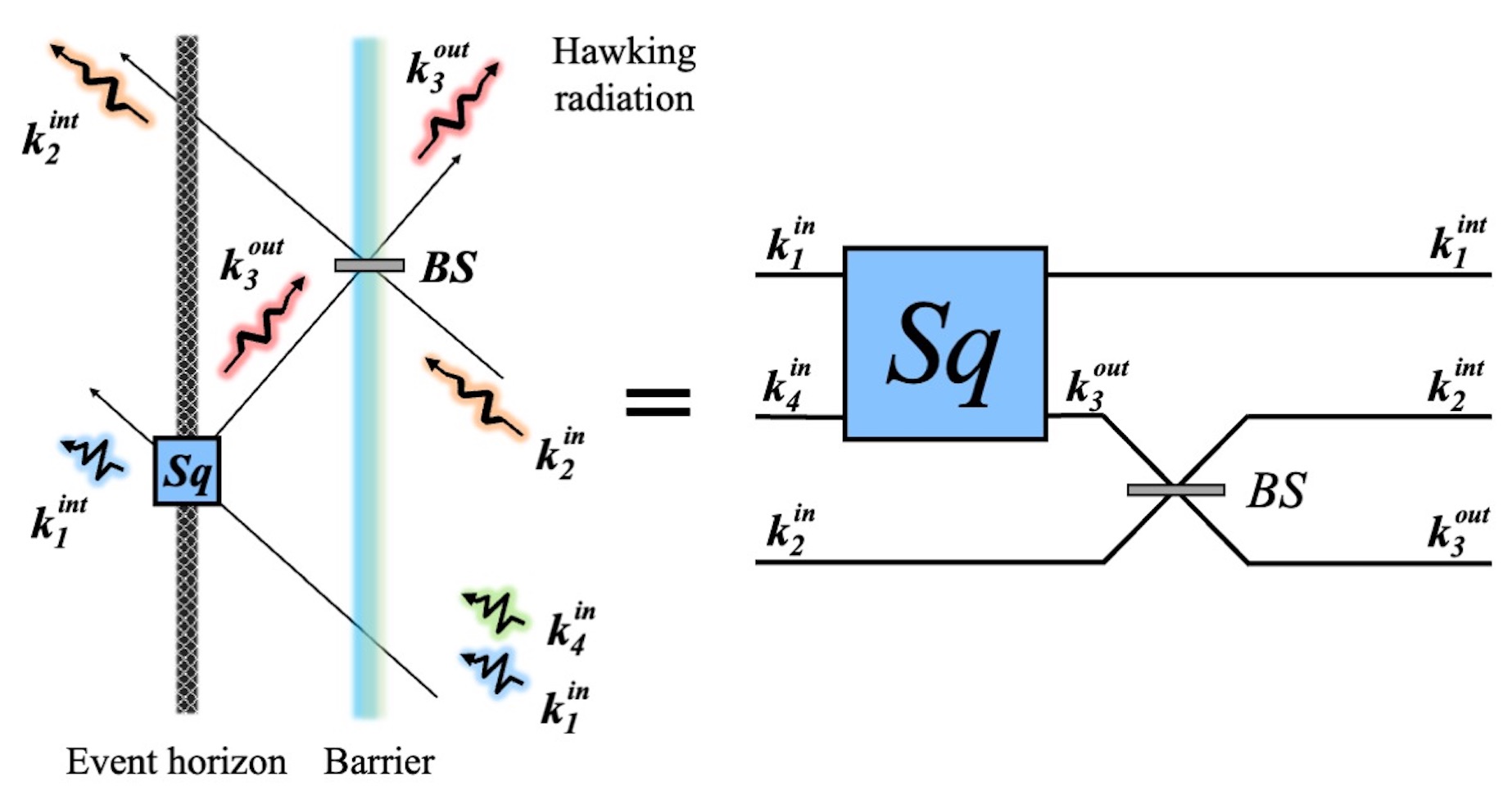}
}
  \caption{Left: Illustration of the two elements responsible for the Hawking process in optical black holes: a two-mode squeezer associated with the horizon, and a beam splitter associated with a process of scattering. Right: Equivalent optical circuit.} 
\label{scattering_v_circuit}
\end{figure}

We propose an analytical expression for the $S$-matrix, obtained by combining elementary operations consisting of two-mode squeezers and beam-splitters, which we choose  by paying  attention to the physics of the problem.
This allows us to map the Hawking process into a simple optical circuit, which helps to visualize the physics and perform computations efficiently.

For pedagogical purposes, we begin by writing an analog optical circuit  exclusively for the black hole side of the pulse, momentarily neglecting the white hole. In the astrophysical case, the evolution  
is dominated by two physical processes, a mixing of positive- and negative-frequency modes induced by the horizon, and a  scattering process due to  the gravitational potential barrier. Mathematically, the first process corresponds to a two-mode squeezer, while the second to a beam-splitter. The situation is analog for optical black holes, except that we have three {\em in} modes, $k^{in}_1$, $k^{in}_2$ and $k^{in}_4$, and three {\em out} modes, $ k^{int}_1$, $ k^{int}_2$ and $k^{out}_3$ (see Fig.~\ref{scattering_v_circuit}). 

It is  straightforward to convert this circuit into an analytic expression for the $S$-matrix. First, recall  the action of a two-mode squeezer on the annihilation operators:  
\bea  a_{ k^{in}_1} &\to& a_{ k^{in}_1}  \cosh{r_H}+ a_{k^{in}_4}^{\dagger}\, e^{i \phi} \, \sinh{r_H}\, ,\nonumber  \\    a_{k^{in}_4}  &\to&  a_{k^{in}_4} \, \cosh{r_H} +   a_{ k^{in}_1}^{\dagger}\, e^{i \phi}\,  \sinh{r_H}\, , \eea 
where $r_H$ and $\phi$ are the intensity and angle, respectively, of the {\em Hawking squeezer}. The action of the beam-splitter is the orthogonal transformation \bea  a_ {k^{in}_2} &\to& a_ {k^{in}_2} \, \cos{\theta}   + a_{k^{out}_3} \sin{\theta} \, ,  \nonumber  \\  a_{ k^{out}_3}&\to& -a_{ k^{in}_2}\,  \sin{\theta}\,  +  a_{k^{out}_3} \cos{\theta} \, , \eea
where $\cos{\theta}$ and $\sin{\theta}$ are the transmission and reflection amplitudes of the splitter, respectively.  Combining these two operations---following the order written in the circuit---and changing variables to the quadrature operators, $x_i\equiv \frac{1}{\sqrt{2}} \, (a_{k_i}+a_{k_i}^{\dagger})$ and $p_i\equiv \frac{-i}{\sqrt{2}} \, (a_{k_i}-a_{k_i}^{\dagger})$, we construct the $S$-matrix  corresponding to the circuit in Fig.~\ref{scattering_v_circuit}, which, when acting on the vector of quadrature operators $\vec r_{in}=(x_1,p_1,x_2,p_2,x_4,p_4)$, implements the Heisenberg evolution: $\vec r_{out}=S\cdot \vec r_{in}$.

With this formalism, it is particularly easy to evolve any Gaussian state, such as  the vacuum, coherent, squeezed, or thermal states. Note that a Gaussian state is completely characterized by its first and second moments, $\vec \mu \equiv \langle \vec r\rangle$ and $\sigma\equiv\langle \{ (\vec r-\vec \mu),  (\vec r-\vec \mu)\}\rangle$, where $\sigma$ is the covariance matrix and  $\{\cdot,\cdot\}$ the symmetric anti-commutator. Because linear evolution preserves  Gaussianity, given an initial Gaussian state ($\vec \mu_{in}$, $\sigma_{in}$), the  final state is also Gaussian, characterized by   $(\vec \mu_{out}, \sigma_{out})=( S\cdot \vec \mu_{in},\  S\cdot \sigma_{in}\cdot S^{\top})$. 

\begin{figure}[t]
{\centering   
\includegraphics[width = 0.45\textwidth]{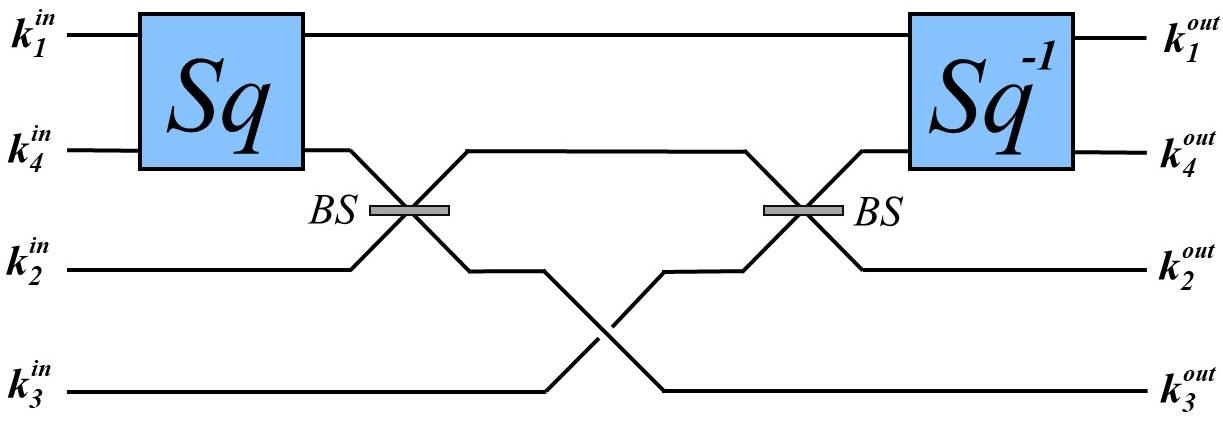}  
}
\caption{White-Black hole analog optical circuit.} 
\label{wbhole_circuit}
\end{figure}

Regarding the white hole, since it is the time reversal of the black hole, its analog optical circuit and corresponding $S$-matrix can be easily obtained by inverting the elements in the black hole circuit. Combining the two, one  obtains the analog circuit for the complete white-black hole system (see Fig.~\ref{wbhole_circuit}). The $S$-matrix for the white-black hole system is then obtained by multiplying the action of squeezers and beam-splitters in the sequence indicated in the circuit. The result is an $8\times 8$ matrix which depends on three parameters: $r_H$, $\theta$, and the phase $\phi$.

{\bf Numerical analysis.} In order to test the accuracy at which our analog circuit describes the physics of the white-black hole system, we have solved the dynamical evolution numerically.  To the best of our knowledge, this is the first complete numerical evolution of waves in a white-black hole system.
We summarize here the most important results of our analysis (a detail description of our numerical infrastructure and its capabilities will appear in  \cite{companion}).

Our code is based on the analytical model proposed in \cite{linder16}, building on previous work \cite{belgiorno15}, and rooted in the Hopfield model, in which the dielectric is made of charged quantum oscillators \cite{hopfield1958}. 
The strong pulse causes a deformation of the energy levels  of these oscillators, which modifies the way the oscillators interact with weak modes, causing a modification of the local refractive index.

Our numerical code solves the dynamical equation in the frequency domain, which is a 4th order ordinary differential equation in the co-moving frame \cite{linder16}. We compute the evolution of {\em in} wave-packets $u^{in}_{k_i}(t_{in})$, with wavenumbers centered on each of the four solutions of the dispersion relation, $(k^{in}_1,k^{in}_2,k^{in}_3,k^{in}_4)$, and with initial spatial support far away from the white-black hole. After  evolving each wave-packet, we decompose the result in the  basis of {\em out} wave-packets centered around $(k^{out}_1,k^{out}_2,k^{out}_3,k^{out}_4)$,
$u^{in}_{k_i}(t_{out})=\sum_j \alpha_{ij}\, u^{out}_{k_j}(t_{out})+\beta_{ij}\,  \bar u^{out}_{k_j}(t_{out})$, where the bar denotes complex conjugation. The Bogoliubov coefficients $\alpha_{ij}$ and $\beta_{ij}$ encode the dynamics, and from these coefficients, we construct the $S$-matrix. 

We model the perturbation of the refractive index as $\delta n(x,t)=\delta n_0\,\sech^2\left(\frac{t-x/u}{\Delta}\right)$ (a common choice in the literature \cite{philbin08, drori19}), where $u$ is the speed of the perturbation; $x$ and $t$ are space-time coordinates in the lab frame; and $\delta n_{0}$ and $\Delta$ determine its amplitude and width, respectively. We treat $\delta n_{0}$ and $\Delta$ as free parameters in what follows. We have performed simulations for $\delta n_{0}$ and $\Delta$, ranging from $0.01$ to $0.1$, and from $2$ fs to $10$ fs, respectively.  We find this is the range for which the analogy with the Hawking effect works better (see below).

\begin{figure}[t]
{\centering     
\includegraphics[width = \linewidth]{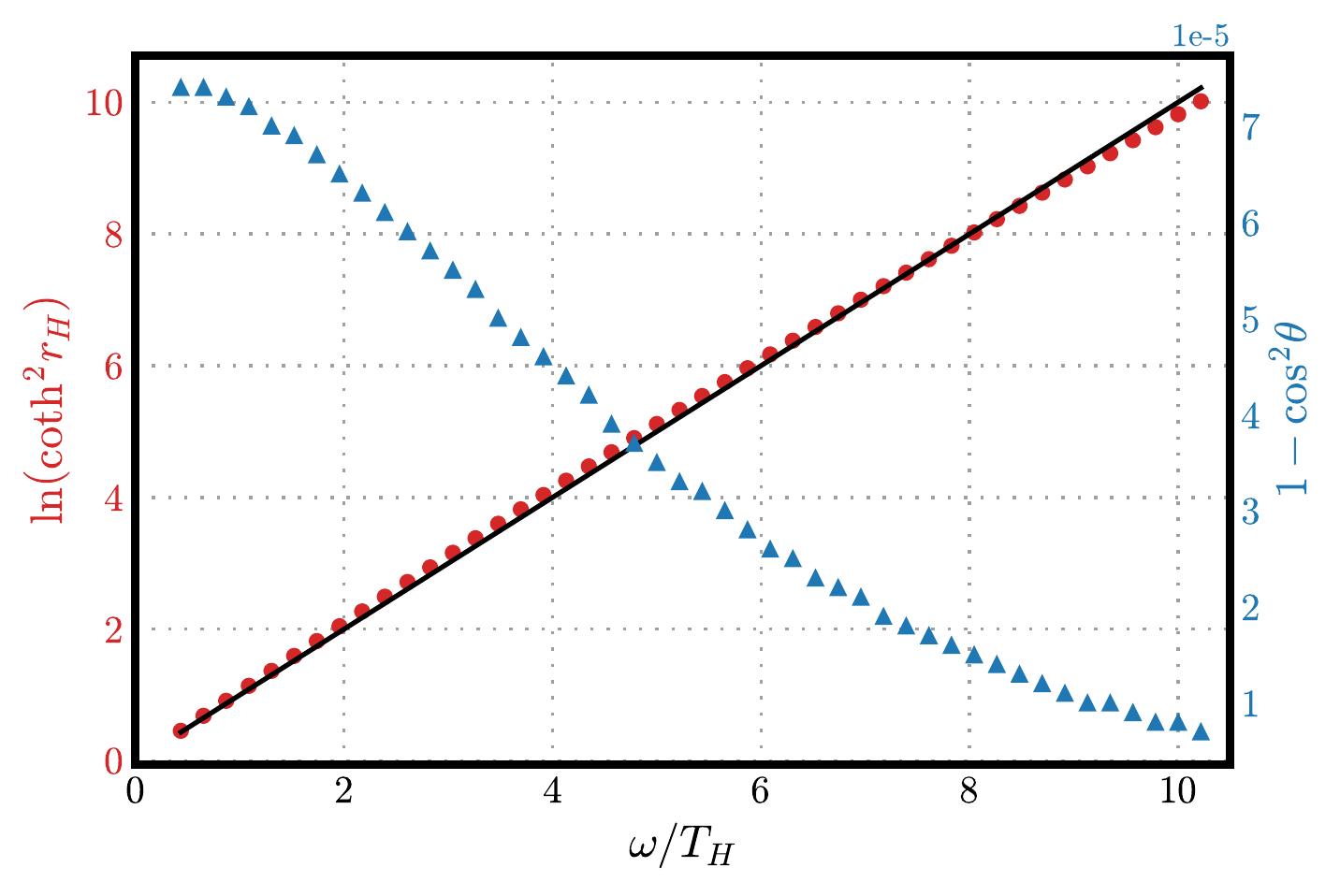}
}
\caption{Left axis (red dots): Numerical results for $\ln(\coth^2r_H)$ vs $\omega$ for a strong pulse determined by $\Delta=6\, fs$ and $\delta n=0.05$. We also show the straight line fit $\omega/T_H$, with $T_H=3.51\, {\rm K}$. Right axis (blue triangles): Deviation of the beamsplitter transmission probability from unity ($1-\cos^2\theta$) vs $\omega$.}
\label{numerics}
\end{figure}

For $\Delta\gtrsim 4$ fs and $0.1 \leq \omega/T_H \leq 5$,
our analytical circuit provides a good approximation for the dynamics, with agreement at the level of (or better than) a percent.
In this regime, we confirm that the intensity of the Hawking squeezer, $r_H$, when computed for different frequencies, follows a Planckian distribution, in the  sense that $\coth^2{r_H}\approx \exp{\frac{\omega}{ T_H}}$ with a temperature $T_H$ which agrees with the analog surface gravity of the horizon at the level of a few percent (see Fig.~\ref{numerics}). (Deviations at the percent level are expected due to dispersive effects.) 
For instance, we find  $T_H=10.4\, {\rm K}$ for $(\Delta=4fs,\delta n_0=0.1$);   $T_H=3.51\, {\rm K}$ for $(\Delta=6fs,\delta n_0=0.05)$; and $T_H=0.52 \, {\rm K}$ for $(\Delta=8fs,\delta n_0=0.01$) for diamond, for which the the refractive index is approximately given by $n^2(\lambda)=1+\frac{4.658\,\lambda^2}{\lambda^2-112.5^2}$ \cite{Diamond}, where  $\lambda$ is the free-space wavelength measured in the lab frame and expressed in nm.

The analysis also reveals interesting subtleties: (i) If the pulse width $\Delta$ and/or the intensity $\delta n_0$ are very small,  
the tunneling probability for the {\em in} mode $k^{in}_3$ to cross the white hole and exit on the black hole side as $k^{out}_3$ becomes non-negligible. This effect is more pronounced for low frequencies. For instance, for $\Delta=2$ fs and $\delta n_0=0.01$,
this effect introduces order-one discrepancies between the numerics and our analytical circuit for $\omega/T_H \lesssim 0.1$, although the discrepancies quickly decrease for larger $\omega$ or larger values of  $\Delta$ and/or $\delta n_0$.  This is an intrinsic limitation of optical analog models, rooted in the fact that right moving modes in the region between the two horizons do actually exist---in contrast to the astrophysical case---although they have exponentially decaying amplitudes. 
(ii) We observe  a  mixing  between   $k^{in}_2$ and $k^{out}_1$ slightly higher than predicted by our circuit. Since these modes have symplectic norms of different signs, this implies that there is another contribution to particle creation, originating from scattering and unrelated to the Hawking process. Such contribution was discussed in a different context in \cite{Corley:1996ar}, although it has not been identified before in optical systems.  We find that this additional particle creation is non-thermal. It impacts the mean number of output quanta in the mode $k_2^{out}$ and its relevance is more important at large frequencies. However, the impact on the most relevant output channels for the Hawking effect---the modes $k^{out}_1,k^{out}_3$ and $k^{out}_4$---is negligible for $\omega/T_H\lesssim 5$ in all our simulations.

{\bf Stimulated Hawking process.} 
We have explored the evolution of a family of Gaussian initial states and have studied the output intensities and entanglement structure generated  during the Hawking process. 
We quantify the entanglement by means of the logarithmic negativity (LogNeg) \cite{vidal02,plenio05}. 
We find that non-classical inputs (squeezed states) alter the covariance of the final state (as opposed to e.g. coherent state inputs) and  can be used to amplify the entanglement generated by the Hawking process. The details depend on the concrete choice of initial squeezed state, and using our formalism we can obtain analytical expressions for all aspects of the out state. 

We have incorporated the effect of losses (e.g. detector inefficiencies) and ambient thermal noise, both ubiquitous in real experiments.  Thermal noise can be incorporated by adding a Planckian  background at temperature $T_{\rm env}$ to all input modes, while  the effects of inefficiencies can be modeled by the following transformation of the final state:  $\vec\mu_{out}\to  \sqrt{\eta}\, \vec\mu_{out}$, $\sigma_{out}\to \eta\,\sigma_{out}+(1-\eta)\, \mathbb{I}$, where $0\leq\eta\leq 1$ is the attenuation factor. We add the same amount of noise and inefficiencies to all channels, although this can be generalized straightforwardly.

\begin{figure}[t]
{\centering     
\includegraphics[width =\linewidth]{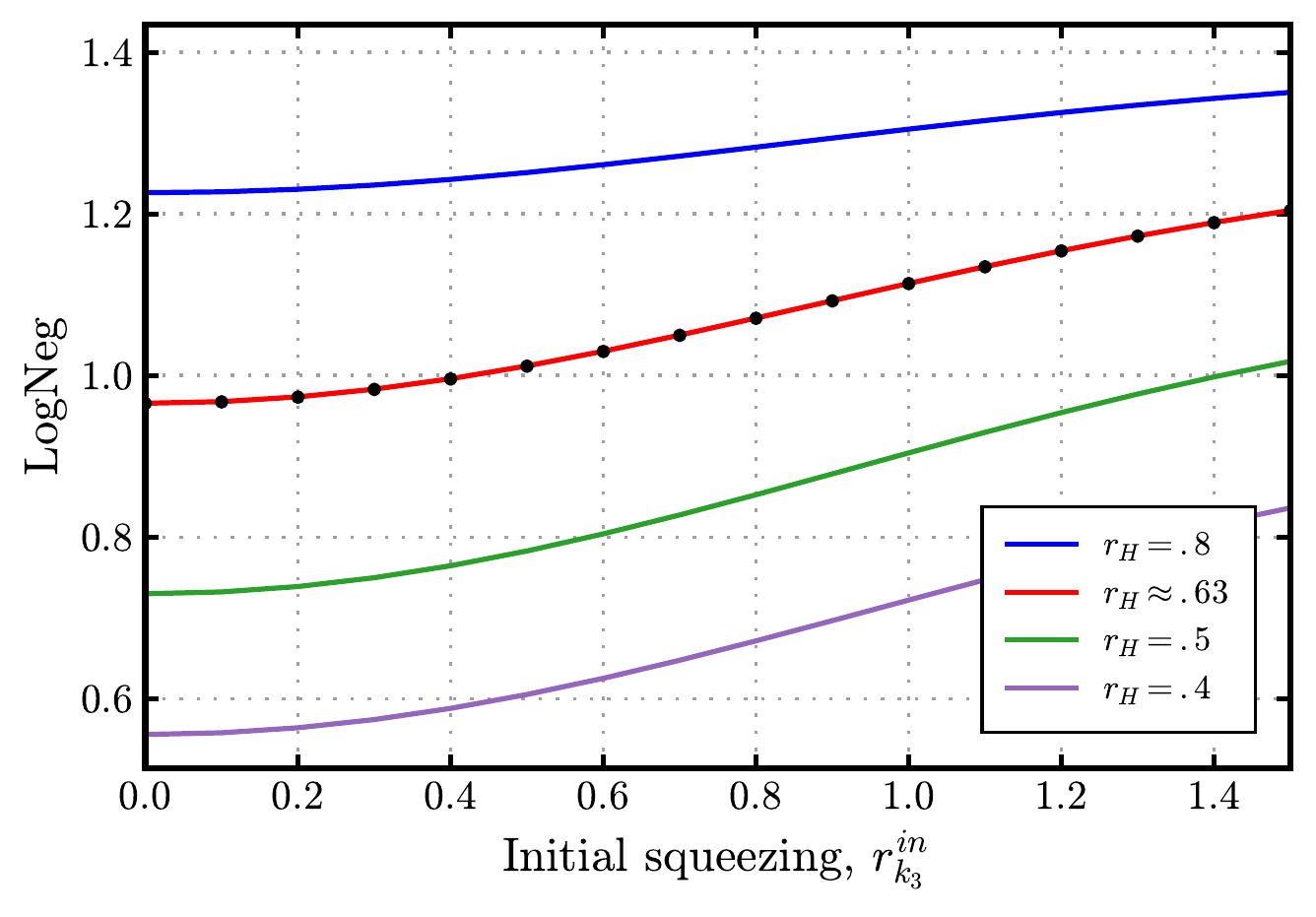} } \caption{Stimulating the white hole for entanglement enhancement. Continuum lines: LogNeg between the outgoing white-hole partner-modes $k^{out}_1$ and $k^{out}_4$ versus the  initial squeezing intensity $r_{k_3}^{in}$, computed from the circuit model for various values of the Hawking squeezing strength, $r_H$. Curves from top to bottom indicate decreasing values of $r_H$. Dots: Results  from the numerical simulations evaluated at $\omega/T_H=1.08$ (same conditions as Fig. \ref{numerics}). Noise parameters are $T_{\rm env}=T_H/2$ and $\eta=0.9$.}
 \label{lognegz}
\end{figure}

{\bf A simple protocol.}  
We find that a convenient strategy is to illuminate the white hole with a single-mode squeezed state in the long wavelength mode $k^{in}_3$, and observe the Hawking-pair of modes $(k_1^{out},k_4^{out})$ leaving the white hole (see Fig.~\ref{modestructure}). This strategy produces an optimal amount of entanglement enhancement, carried by the $(k_1^{out},k_4^{out})$ mode pair, and more importantly, it allows us to recover the information about the Hawking process in a simple manner, as we now describe. 

From the optical circuit, we find that the mean particle number $\langle n^{out}_{k_i}\rangle$ for $i=1,4$ grows {\em linearly} with $\sinh^2{r^{in}_{k_3}}$, where $r^{in}_{k_3}$ is the squeezing intensity chosen for the input state. The rates of these linear growths are given by
\bea \label{ints} m_{k^{out}_1}&=&\eta\, (1 + 2\,  n_{\rm env})\, \cos^2\theta \, \sinh^2r_H\, , \nonumber \\ 
m_{k^{out}_4}&=&\eta\, (1 + 2 \, n_{\rm env})\, \cos^2\theta \, \cosh^2r_H \, , \label{rates} \eea
where $n_{env}$ is the mean number of thermal photons in the environment. 
These rates can be determined in the lab  by measuring the intensity of the output modes and tuning the initial squeezing $r^{in}_{k_3}$. By taking ratios, one can obtain the intensity of the Hawking squeezer $r_H(\omega)$   as $m_{k^{out}_4}/m_{k^{out}_1}=\coth^2r_H$.  
The effects of thermal noise and inefficiencies cancel out in this ratio.  

Although this protocol permits one to reconstruct the properties of the Hawking squeezer, $r_H(\omega)$, it is based on intensities and does not involve any genuinely quantum property. Interestingly, the $r_H(\omega)$ can be independently reconstructed from the entanglement (LogNeg) between the Hawking-pair $(k^{out}_1,k^{out}_4)$ emitted by the white hole. The analytical expression for the LogNeg between these two modes is lengthy, and its behavior with the initial squeezing intensity $r^{in}_{k_3}$ is better illustrated in Fig.~\ref{lognegz}. There are two important takeaway messages from our analysis: (i) In the absence of the Hawking squeezer, $r_H(\omega)=0$, there is no entanglement between $k^{out}_1$, $k^{out}_4$ (not explicitly shown in Fig.~\ref{lognegz}), no matter what the value of the initial single-mode squeezing $r^{in}_{k_3}$ is. Therefore, the observation of such entanglement must be attributed to the Hawking effect, and not to the initial state, which contains no entanglement between these two modes. (ii) The LogNeg increases monotonically with the initial squeezing intensity $r^{in}_{k_3}$ (if inefficiencies are small; see below), and thus initial squeezing enhances the quantum properties of the output. 
Obtaining the LogNeg, for instance by reduced-state reconstruction using homodyne measurements \cite{raymer2009}, and comparing with the theoretical curves in Fig.~\ref{lognegz}, the Hawking squeezing strength, $r_H(\omega)$, can be obtained from a quantity of purely quantum origin. 
The value of $r_H(\omega)$ obtained in this way must agree with the one independently obtained from intensities [Eqns.~(\ref{ints})], providing a strong consistency test. 

To illustrate the way this protocol works, we have added to Fig.~\ref{lognegz} the results for the LogNeg obtained from numerical simulation, for different initial squeezing $r^{in}_{k_3}$. The numerical simulations are completely independent from our analytical calculations.   By comparing with a family of theoretical curves obtained for different $r_H$,  we
can identify the value of the Hawking squeezing intensity $r_H(\omega)$ which corresponds to the numerical simulations. In a real experiment one would proceed in a similar fashion, by replacing the numerically generated points in Fig.~\ref{lognegz} with experimental data.

We now discuss the effects of thermal noise and inefficiencies.  Independently of  the latter, thermal noise systematically reduces the amount of entanglement in the final state, even causing the entanglement to vanish if the environment temperature  $T_{\rm env}$ is large enough. For instance, for vacuum input, the entanglement in the reduced bipartition $(k^{out}_1| k^{out}_4)$  disappears when $T_{\rm env}$ is larger than a frequency dependent threshold, which we find to be equal to $T_H$ for low frequencies ($\omega/T_H\ll 1$) and equal to $2 \, T_H$ for large frequencies ($\omega/T_H\gg 1$). This last result is in agreement with previous findings in \cite{bruschi2013}. For the bipartition  $(k^{out}_1|k^{out}_3,k^{out}_4)$ these thresholds become $4\, T_H^2/\omega$ and $2T_H/(1-\ln(2)T_H/ \omega)$, respectively. In the absence of inefficiencies ($\eta\approx 1$), squeezing a single mode in the initial state can {\em always} be used to overcome these thresholds and restore the entanglement. 

On the other hand, entanglement is sensitive to the effects of inefficiencies even when initial squeezing is present. In particular, for values of the attenuation parameter $\eta$ smaller than a critical value  $\eta_c$, the effect of squeezing the input is reversed, and initial squeezing {\em degrades} the entanglement in the output. 
For instance, we find $\eta_c\approx 0.8$ for $\omega/T_H=1$, while $\eta_c\approx 0.9$ for $\omega/ T_H=4$.

{\bf Conclusions.}  
This work  introduces an analytical description of the Hawking process on analog optical white-black holes, based on techniques from Gaussian quantum optics. This analysis is complemented with detailed numerical simulations, to assert the accuracy and to determine the regime of applicability of the analog Hawking process. We have analyzed the effects of background noise and inefficiencies and found that they are enemies to the observability  of quantum aspects of  the Hawking process--- i.e. the quantum entanglement between different output channels is easily masked, or completely erased, by these deleterious effects. We have thus been able to quantify the boundary in the parameter space where the observability of quantum aspects of the Hawking process is possible. 

Furthermore, we have introduced a strategy to amplify the quantum features and overcome entanglement-degrading effects and have proposed a protocol to observe them in the lab. Although additional difficulties may arise in a real experiment, our ideas constitute a step forward in the observability of the Hawking process. We have focused on optical systems, but our ideas can be extended to other analog models as well. Additional details and ramifications of our work will be the goal of a companion publication. \\

{\bf Acknowledgments.} We have benefited from  discussions with A. Fabbri, J. Olmedo, J. Pullin,  and specially with O. Magana-Loaiza. We thank J. Olmedo for assistance with Fig~\ref{numerics}. This work is supported by the NSF grants CAREER PHY-1552603 and PHY-2110273, and from the Hearne Institute for Theoretical Physics.

\bibliography{agrav.bib}

\begin{thebibliography}{31}%
\makeatletter
\providecommand \@ifxundefined [1]{%
 \@ifx{#1\undefined}
}%
\providecommand \@ifnum [1]{%
 \ifnum #1\expandafter \@firstoftwo
 \else \expandafter \@secondoftwo
 \fi
}%
\providecommand \@ifx [1]{%
 \ifx #1\expandafter \@firstoftwo
 \else \expandafter \@secondoftwo
 \fi
}%
\providecommand \natexlab [1]{#1}%
\providecommand \enquote  [1]{``#1''}%
\providecommand \bibnamefont  [1]{#1}%
\providecommand \bibfnamefont [1]{#1}%
\providecommand \citenamefont [1]{#1}%
\providecommand \href@noop [0]{\@secondoftwo}%
\providecommand \href [0]{\begingroup \@sanitize@url \@href}%
\providecommand \@href[1]{\@@startlink{#1}\@@href}%
\providecommand \@@href[1]{\endgroup#1\@@endlink}%
\providecommand \@sanitize@url [0]{\catcode `\\12\catcode `\$12\catcode
  `\&12\catcode `\#12\catcode `\^12\catcode `\_12\catcode `\%12\relax}%
\providecommand \@@startlink[1]{}%
\providecommand \@@endlink[0]{}%
\providecommand \url  [0]{\begingroup\@sanitize@url \@url }%
\providecommand \@url [1]{\endgroup\@href {#1}{\urlprefix }}%
\providecommand \urlprefix  [0]{URL }%
\providecommand \Eprint [0]{\href }%
\providecommand \doibase [0]{https://doi.org/}%
\providecommand \selectlanguage [0]{\@gobble}%
\providecommand \bibinfo  [0]{\@secondoftwo}%
\providecommand \bibfield  [0]{\@secondoftwo}%
\providecommand \translation [1]{[#1]}%
\providecommand \BibitemOpen [0]{}%
\providecommand \bibitemStop [0]{}%
\providecommand \bibitemNoStop [0]{.\EOS\space}%
\providecommand \EOS [0]{\spacefactor3000\relax}%
\providecommand \BibitemShut  [1]{\csname bibitem#1\endcsname}%
\let\auto@bib@innerbib\@empty
%</preamble>
\bibitem [{\citenamefont {Hawking}(1974)}]{Hawking:1974rv}%
  \BibitemOpen
  \bibfield  {author} {\bibinfo {author} {\bibfnamefont {S.~W.}\ \bibnamefont
  {Hawking}},\ }\bibfield  {title} {\bibinfo {title} {{Black hole
  explosions}},\ }\href {https://doi.org/10.1038/248030a0} {\bibfield
  {journal} {\bibinfo  {journal} {Nature}\ }\textbf {\bibinfo {volume} {248}},\
  \bibinfo {pages} {30} (\bibinfo {year} {1974})}\BibitemShut {NoStop}%
\bibitem [{\citenamefont {Hawking}(1975)}]{Hawking:1975vcx}%
  \BibitemOpen
  \bibfield  {author} {\bibinfo {author} {\bibfnamefont {S.~W.}\ \bibnamefont
  {Hawking}},\ }\bibfield  {title} {\bibinfo {title} {{Particle Creation by
  Black Holes}},\ }\href {https://doi.org/10.1007/BF02345020} {\bibfield
  {journal} {\bibinfo  {journal} {Commun. Math. Phys.}\ }\textbf {\bibinfo
  {volume} {43}},\ \bibinfo {pages} {199} (\bibinfo {year} {1975})},\ \bibinfo
  {note} {[Erratum: Commun.Math.Phys. 46, 206 (1976)]}\BibitemShut {NoStop}%
\bibitem [{\citenamefont {Bardeen}\ \emph {et~al.}(1973)\citenamefont
  {Bardeen}, \citenamefont {Carter},\ and\ \citenamefont
  {Hawking}}]{Bardeen:1973gs}%
  \BibitemOpen
  \bibfield  {author} {\bibinfo {author} {\bibfnamefont {J.~M.}\ \bibnamefont
  {Bardeen}}, \bibinfo {author} {\bibfnamefont {B.}~\bibnamefont {Carter}},\
  and\ \bibinfo {author} {\bibfnamefont {S.~W.}\ \bibnamefont {Hawking}},\
  }\bibfield  {title} {\bibinfo {title} {{The Four laws of black hole
  mechanics}},\ }\href {https://doi.org/10.1007/BF01645742} {\bibfield
  {journal} {\bibinfo  {journal} {Commun. Math. Phys.}\ }\textbf {\bibinfo
  {volume} {31}},\ \bibinfo {pages} {161} (\bibinfo {year} {1973})}\BibitemShut
  {NoStop}%
\bibitem [{\citenamefont {Bekenstein}(1974)}]{Bekenstein:1974ax}%
  \BibitemOpen
  \bibfield  {author} {\bibinfo {author} {\bibfnamefont {J.~D.}\ \bibnamefont
  {Bekenstein}},\ }\bibfield  {title} {\bibinfo {title} {{Generalized second
  law of thermodynamics in black hole physics}},\ }\href
  {https://doi.org/10.1103/PhysRevD.9.3292} {\bibfield  {journal} {\bibinfo
  {journal} {Phys. Rev. D}\ }\textbf {\bibinfo {volume} {9}},\ \bibinfo {pages}
  {3292} (\bibinfo {year} {1974})}\BibitemShut {NoStop}%
\bibitem [{\citenamefont {{Visser, Matt}}(2003)}]{Visser2003}%
  \BibitemOpen
  \bibfield  {author} {\bibinfo {author} {\bibnamefont {{Visser, Matt}}},\
  }\bibfield  {title} {\bibinfo {title} {{Essential and inessential features of
  Hawking radiation}},\ }\href {https://doi.org/10.1142/S0218271803003190}
  {\bibfield  {journal} {\bibinfo  {journal} {International Journal of Modern
  Physics D}\ }\textbf {\bibinfo {volume} {12}},\ \bibinfo {pages} {649}
  (\bibinfo {year} {2003})}\BibitemShut {NoStop}%
\bibitem [{\citenamefont {Jacobson}(1991)}]{Jacobson:1991gr}%
  \BibitemOpen
  \bibfield  {author} {\bibinfo {author} {\bibfnamefont {T.}~\bibnamefont
  {Jacobson}},\ }\bibfield  {title} {\bibinfo {title} {{Black hole evaporation
  and ultrashort distances}},\ }\href
  {https://doi.org/10.1103/PhysRevD.44.1731} {\bibfield  {journal} {\bibinfo
  {journal} {Phys. Rev. D}\ }\textbf {\bibinfo {volume} {44}},\ \bibinfo
  {pages} {1731} (\bibinfo {year} {1991})}\BibitemShut {NoStop}%
\bibitem [{\citenamefont {Hawking}(1976)}]{Hawking:1976ra}%
  \BibitemOpen
  \bibfield  {author} {\bibinfo {author} {\bibfnamefont {S.~W.}\ \bibnamefont
  {Hawking}},\ }\bibfield  {title} {\bibinfo {title} {{Breakdown of
  Predictability in Gravitational Collapse}},\ }\href
  {https://doi.org/10.1103/PhysRevD.14.2460} {\bibfield  {journal} {\bibinfo
  {journal} {Phys. Rev. D}\ }\textbf {\bibinfo {volume} {14}},\ \bibinfo
  {pages} {2460} (\bibinfo {year} {1976})}\BibitemShut {NoStop}%
\bibitem [{\citenamefont {{Philbin, Thomas G and Kuklewicz, Chris and
  Robertson, Scott and Hill, Stephen and K{\"o}nig, Friedrich and Leonhardt,
  Ulf}}(2008)}]{philbin08}%
  \BibitemOpen
  \bibfield  {author} {\bibinfo {author} {\bibnamefont {{Philbin, Thomas G and
  Kuklewicz, Chris and Robertson, Scott and Hill, Stephen and K{\"o}nig,
  Friedrich and Leonhardt, Ulf}}},\ }\bibfield  {title} {\bibinfo {title}
  {Fiber-optical analog of the event horizon},\ }\href
  {https://doi.org/10.1126/science.1153625} {\bibfield  {journal} {\bibinfo
  {journal} {Science}\ }\textbf {\bibinfo {volume} {319}},\ \bibinfo {pages}
  {1367} (\bibinfo {year} {2008})}\BibitemShut {NoStop}%
\bibitem [{\citenamefont {{Weinfurtner, Silke and Tedford, Edmund W and
  Penrice, Matthew CJ and Unruh, William G and Lawrence, Gregory
  A}}(2011)}]{weinfurtner2011}%
  \BibitemOpen
  \bibfield  {author} {\bibinfo {author} {\bibnamefont {{Weinfurtner, Silke and
  Tedford, Edmund W and Penrice, Matthew CJ and Unruh, William G and Lawrence,
  Gregory A}}},\ }\bibfield  {title} {\bibinfo {title} {{Measurement of
  stimulated Hawking emission in an analogue system}},\ }\href
  {https://doi.org/10.1103/PhysRevLett.106.021302} {\bibfield  {journal}
  {\bibinfo  {journal} {Physical Review Letters}\ }\textbf {\bibinfo {volume}
  {106}},\ \bibinfo {pages} {021302} (\bibinfo {year} {2011})}\BibitemShut
  {NoStop}%
\bibitem [{\citenamefont {Euv{\'e}}\ \emph {et~al.}(2016)\citenamefont
  {Euv{\'e}}, \citenamefont {Michel}, \citenamefont {Parentani}, \citenamefont
  {Philbin},\ and\ \citenamefont {Rousseaux}}]{euve2016}%
  \BibitemOpen
  \bibfield  {author} {\bibinfo {author} {\bibfnamefont {L.-P.}\ \bibnamefont
  {Euv{\'e}}}, \bibinfo {author} {\bibfnamefont {F.}~\bibnamefont {Michel}},
  \bibinfo {author} {\bibfnamefont {R.}~\bibnamefont {Parentani}}, \bibinfo
  {author} {\bibfnamefont {T.~G.}\ \bibnamefont {Philbin}},\ and\ \bibinfo
  {author} {\bibfnamefont {G.}~\bibnamefont {Rousseaux}},\ }\bibfield  {title}
  {\bibinfo {title} {{Observation of noise correlated by the Hawking effect in
  a water tank}},\ }\href {https://doi.org/10.1103/PhysRevLett.117.121301}
  {\bibfield  {journal} {\bibinfo  {journal} {Physical Review Letters}\
  }\textbf {\bibinfo {volume} {117}},\ \bibinfo {pages} {121301} (\bibinfo
  {year} {2016})}\BibitemShut {NoStop}%
\bibitem [{\citenamefont {{Steinhauer, Jeff}}(2016)}]{steinhauer2016}%
  \BibitemOpen
  \bibfield  {author} {\bibinfo {author} {\bibnamefont {{Steinhauer, Jeff}}},\
  }\bibfield  {title} {\bibinfo {title} {{Observation of quantum Hawking
  radiation and its entanglement in an analogue black hole}},\ }\href
  {https://doi.org/10.1038/nphys3863} {\bibfield  {journal} {\bibinfo
  {journal} {Nature Physics}\ }\textbf {\bibinfo {volume} {12}},\ \bibinfo
  {pages} {959} (\bibinfo {year} {2016})}\BibitemShut {NoStop}%
\bibitem [{\citenamefont {{De Nova, Juan Ramon Munoz and Golubkov, Katrine and
  Kolobov, Victor I and Steinhauer, Jeff}}(2019)}]{de19BEC}%
  \BibitemOpen
  \bibfield  {author} {\bibinfo {author} {\bibnamefont {{De Nova, Juan Ramon
  Munoz and Golubkov, Katrine and Kolobov, Victor I and Steinhauer, Jeff}}},\
  }\bibfield  {title} {\bibinfo {title} {{Observation of thermal Hawking
  radiation and its temperature in an analogue black hole}},\ }\href
  {https://doi.org/10.1038/s41586-019-1241-0} {\bibfield  {journal} {\bibinfo
  {journal} {Nature}\ }\textbf {\bibinfo {volume} {569}},\ \bibinfo {pages}
  {688} (\bibinfo {year} {2019})}\BibitemShut {NoStop}%
\bibitem [{\citenamefont {{Drori, Jonathan and Rosenberg, Yuval and Bermudez,
  David and Silberberg, Yaron and Leonhardt, Ulf}}(2019)}]{drori19}%
  \BibitemOpen
  \bibfield  {author} {\bibinfo {author} {\bibnamefont {{Drori, Jonathan and
  Rosenberg, Yuval and Bermudez, David and Silberberg, Yaron and Leonhardt,
  Ulf}}},\ }\bibfield  {title} {\bibinfo {title} {{Observation of stimulated
  Hawking radiation in an optical analogue}},\ }\href
  {https://doi.org/10.1103/PhysRevLett.122.010404} {\bibfield  {journal}
  {\bibinfo  {journal} {Physical Review Letters}\ }\textbf {\bibinfo {volume}
  {122}},\ \bibinfo {pages} {010404} (\bibinfo {year} {2019})}\BibitemShut
  {NoStop}%
\bibitem [{\citenamefont {{Kolobov, Victor I and Golubkov, Katrine and de Nova,
  Juan Ram{\'o}n Mu{\~n}oz and Steinhauer, Jeff}}(2021)}]{kolobov2021BEC}%
  \BibitemOpen
  \bibfield  {author} {\bibinfo {author} {\bibnamefont {{Kolobov, Victor I and
  Golubkov, Katrine and de Nova, Juan Ram{\'o}n Mu{\~n}oz and Steinhauer,
  Jeff}}},\ }\bibfield  {title} {\bibinfo {title} {{Observation of stationary
  spontaneous Hawking radiation and the time evolution of an analogue black
  hole}},\ }\href {https://doi.org/10.1038/s41567-020-01076-0} {\bibfield
  {journal} {\bibinfo  {journal} {Nature Physics}\ ,\ \bibinfo {pages} {1}}
  (\bibinfo {year} {2021})}\BibitemShut {NoStop}%
\bibitem [{\citenamefont {Lvovsky}\ and\ \citenamefont
  {Raymer}(2009)}]{raymer2009}%
  \BibitemOpen
  \bibfield  {author} {\bibinfo {author} {\bibfnamefont {A.~I.}\ \bibnamefont
  {Lvovsky}}\ and\ \bibinfo {author} {\bibfnamefont {M.~G.}\ \bibnamefont
  {Raymer}},\ }\bibfield  {title} {\bibinfo {title} {Continuous-variable
  optical quantum-state tomography},\ }\href
  {https://doi.org/10.1103/RevModPhys.81.299} {\bibfield  {journal} {\bibinfo
  {journal} {Reviews of Modern Physics}\ }\textbf {\bibinfo {volume} {81}},\
  \bibinfo {pages} {299} (\bibinfo {year} {2009})}\BibitemShut {NoStop}%
\bibitem [{\citenamefont {{Demircan, A and Amiranashvili, Sh and Steinmeyer,
  G}}(2011)}]{demircan11TRANSISTOR}%
  \BibitemOpen
  \bibfield  {author} {\bibinfo {author} {\bibnamefont {{Demircan, A and
  Amiranashvili, Sh and Steinmeyer, G}}},\ }\bibfield  {title} {\bibinfo
  {title} {Controlling light by light with an optical event horizon},\ }\href
  {https://doi.org/10.1103/PhysRevLett.106.163901} {\bibfield  {journal}
  {\bibinfo  {journal} {Physical Review Letters}\ }\textbf {\bibinfo {volume}
  {106}},\ \bibinfo {pages} {163901} (\bibinfo {year} {2011})}\BibitemShut
  {NoStop}%
\bibitem [{\citenamefont {{Rubino, Elenora and Lotti, A and Belgiorno, F and
  Cacciatori, SL and Couairon, Arnaud and Leonhardt, Ulf and Faccio,
  D}}(2012)}]{rubino2012soliton}%
  \BibitemOpen
  \bibfield  {author} {\bibinfo {author} {\bibnamefont {{Rubino, Elenora and
  Lotti, A and Belgiorno, F and Cacciatori, SL and Couairon, Arnaud and
  Leonhardt, Ulf and Faccio, D}}},\ }\bibfield  {title} {\bibinfo {title}
  {Soliton-induced relativistic-scattering and amplification},\ }\href
  {https://doi.org/10.1038/srep00932} {\bibfield  {journal} {\bibinfo
  {journal} {Scientific Reports}\ }\textbf {\bibinfo {volume} {2}},\ \bibinfo
  {pages} {1} (\bibinfo {year} {2012})}\BibitemShut {NoStop}%
\bibitem [{\citenamefont {{Petev, Mike and Westerberg, Niclas and Moss, Daniel
  and Rubino, Elenora and Rimoldi, C and Cacciatori, SL and Belgiorno, F and
  Faccio, D}}(2013)}]{petev2013blackbody}%
  \BibitemOpen
  \bibfield  {author} {\bibinfo {author} {\bibnamefont {{Petev, Mike and
  Westerberg, Niclas and Moss, Daniel and Rubino, Elenora and Rimoldi, C and
  Cacciatori, SL and Belgiorno, F and Faccio, D}}},\ }\bibfield  {title}
  {\bibinfo {title} {Blackbody emission from light interacting with an
  effective moving dispersive medium},\ }\href
  {https://doi.org/10.1103/PhysRevLett.111.043902} {\bibfield  {journal}
  {\bibinfo  {journal} {Physical Review Letters}\ }\textbf {\bibinfo {volume}
  {111}},\ \bibinfo {pages} {043902} (\bibinfo {year} {2013})}\BibitemShut
  {NoStop}%
\bibitem [{\citenamefont {{Rosenberg, Yuval}}(2020)}]{rosenberg2020optical}%
  \BibitemOpen
  \bibfield  {author} {\bibinfo {author} {\bibnamefont {{Rosenberg, Yuval}}},\
  }\bibfield  {title} {\bibinfo {title} {Optical analogues of black-hole
  horizons},\ }\href {https://doi.org/10.1098/rsta.2019.0232} {\bibfield
  {journal} {\bibinfo  {journal} {Philosophical Transactions of the Royal
  Society A}\ }\textbf {\bibinfo {volume} {378}},\ \bibinfo {pages} {20190232}
  (\bibinfo {year} {2020})}\BibitemShut {NoStop}%
\bibitem [{\citenamefont {{Linder, Malte F and Sch{\"u}tzhold, Ralf and Unruh,
  William G}}(2016)}]{linder16}%
  \BibitemOpen
  \bibfield  {author} {\bibinfo {author} {\bibnamefont {{Linder, Malte F and
  Sch{\"u}tzhold, Ralf and Unruh, William G}}},\ }\bibfield  {title} {\bibinfo
  {title} {{Derivation of Hawking radiation in dispersive dielectric media}},\
  }\href {https://doi.org/10.1103/PhysRevD.93.104010} {\bibfield  {journal}
  {\bibinfo  {journal} {Physical Review D}\ }\textbf {\bibinfo {volume} {93}},\
  \bibinfo {pages} {104010} (\bibinfo {year} {2016})}\BibitemShut {NoStop}%
\bibitem [{\citenamefont {Agullo}\ \emph {et~al.}()\citenamefont {Agullo},
  \citenamefont {Brady},\ and\ \citenamefont {Kranas}}]{companion}%
  \BibitemOpen
  \bibfield  {author} {\bibinfo {author} {\bibfnamefont {I.}~\bibnamefont
  {Agullo}}, \bibinfo {author} {\bibfnamefont {A.}~\bibnamefont {Brady}},\ and\
  \bibinfo {author} {\bibfnamefont {D.}~\bibnamefont {Kranas}},\ }\bibfield
  {title} {\bibinfo {title} {{To appear}},\ }\href@noop {} {\ }\BibitemShut
  {NoStop}%
\bibitem [{\citenamefont {{Belgiorno, F and Cacciatori, SL and Dalla Piazza,
  F}}(2015)}]{belgiorno15}%
  \BibitemOpen
  \bibfield  {author} {\bibinfo {author} {\bibnamefont {{Belgiorno, F and
  Cacciatori, SL and Dalla Piazza, F}}},\ }\bibfield  {title} {\bibinfo {title}
  {{Hawking effect in dielectric media and the Hopfield model}},\ }\href
  {https://doi.org/10.1103/PhysRevD.91.124063} {\bibfield  {journal} {\bibinfo
  {journal} {Physical Review D}\ }\textbf {\bibinfo {volume} {91}},\ \bibinfo
  {pages} {124063} (\bibinfo {year} {2015})}\BibitemShut {NoStop}%
\bibitem [{\citenamefont {Hopfield}(1958)}]{hopfield1958}%
  \BibitemOpen
  \bibfield  {author} {\bibinfo {author} {\bibfnamefont {J.}~\bibnamefont
  {Hopfield}},\ }\bibfield  {title} {\bibinfo {title} {Theory of the
  contribution of excitons to the complex dielectric constant of crystals},\
  }\href {https://doi.org/10.1103/PhysRev.112.1555} {\bibfield  {journal}
  {\bibinfo  {journal} {Physical Review}\ }\textbf {\bibinfo {volume} {112}},\
  \bibinfo {pages} {1555} (\bibinfo {year} {1958})}\BibitemShut {NoStop}%
\bibitem [{\citenamefont {Turri}\ \emph {et~al.}(2017)\citenamefont {Turri},
  \citenamefont {Webster}, \citenamefont {Chen}, \citenamefont {Wickham},
  \citenamefont {Bennett},\ and\ \citenamefont {Bass}}]{Diamond}%
  \BibitemOpen
  \bibfield  {author} {\bibinfo {author} {\bibfnamefont {G.}~\bibnamefont
  {Turri}}, \bibinfo {author} {\bibfnamefont {S.}~\bibnamefont {Webster}},
  \bibinfo {author} {\bibfnamefont {Y.}~\bibnamefont {Chen}}, \bibinfo {author}
  {\bibfnamefont {B.}~\bibnamefont {Wickham}}, \bibinfo {author} {\bibfnamefont
  {A.}~\bibnamefont {Bennett}},\ and\ \bibinfo {author} {\bibfnamefont
  {M.}~\bibnamefont {Bass}},\ }\bibfield  {title} {\bibinfo {title} {Index of
  refraction from the near-ultraviolet to the near-infrared from a single
  crystal microwave-assisted cvd diamond},\ }\href
  {https://doi.org/10.1364/OME.7.000855} {\bibfield  {journal} {\bibinfo
  {journal} {Opt. Mater. Express}\ }\textbf {\bibinfo {volume} {7}},\ \bibinfo
  {pages} {855} (\bibinfo {year} {2017})}\BibitemShut {NoStop}%
\bibitem [{\citenamefont {Corley}\ and\ \citenamefont
  {Jacobson}(1996)}]{Corley:1996ar}%
  \BibitemOpen
  \bibfield  {author} {\bibinfo {author} {\bibfnamefont {S.}~\bibnamefont
  {Corley}}\ and\ \bibinfo {author} {\bibfnamefont {T.}~\bibnamefont
  {Jacobson}},\ }\bibfield  {title} {\bibinfo {title} {{Hawking spectrum and
  high frequency dispersion}},\ }\href
  {https://doi.org/10.1103/PhysRevD.54.1568} {\bibfield  {journal} {\bibinfo
  {journal} {Phys. Rev. D}\ }\textbf {\bibinfo {volume} {54}},\ \bibinfo
  {pages} {1568} (\bibinfo {year} {1996})},\ \Eprint
  {https://arxiv.org/abs/hep-th/9601073} {arXiv:hep-th/9601073} \BibitemShut
  {NoStop}%
\bibitem [{\citenamefont {Vidal}\ and\ \citenamefont {Werner}(2002)}]{vidal02}%
  \BibitemOpen
  \bibfield  {author} {\bibinfo {author} {\bibfnamefont {G.}~\bibnamefont
  {Vidal}}\ and\ \bibinfo {author} {\bibfnamefont {R.~F.}\ \bibnamefont
  {Werner}},\ }\bibfield  {title} {\bibinfo {title} {Computable measure of
  entanglement},\ }\href {https://doi.org/10.1103/PhysRevA.65.032314}
  {\bibfield  {journal} {\bibinfo  {journal} {Physical Review A}\ }\textbf
  {\bibinfo {volume} {65}},\ \bibinfo {pages} {032314} (\bibinfo {year}
  {2002})}\BibitemShut {NoStop}%
\bibitem [{\citenamefont {Plenio}(2005)}]{plenio05}%
  \BibitemOpen
  \bibfield  {author} {\bibinfo {author} {\bibfnamefont {M.~B.}\ \bibnamefont
  {Plenio}},\ }\bibfield  {title} {\bibinfo {title} {Logarithmic negativity: A
  full entanglement monotone that is not convex},\ }\href
  {https://doi.org/10.1103/PhysRevLett.95.090503} {\bibfield  {journal}
  {\bibinfo  {journal} {Physical Review Letters}\ }\textbf {\bibinfo {volume}
  {95}},\ \bibinfo {pages} {090503} (\bibinfo {year} {2005})}\BibitemShut
  {NoStop}%
\bibitem [{\citenamefont {{Bruschi, David Edward and Friis, Nicolai and
  Fuentes, Ivette and Weinfurtner, Silke}}(2013)}]{bruschi2013}%
  \BibitemOpen
  \bibfield  {author} {\bibinfo {author} {\bibnamefont {{Bruschi, David Edward
  and Friis, Nicolai and Fuentes, Ivette and Weinfurtner, Silke}}},\ }\bibfield
   {title} {\bibinfo {title} {On the robustness of entanglement in analogue
  gravity systems},\ }\href {https://doi.org/10.1088/1367-2630/15/11/113016}
  {\bibfield  {journal} {\bibinfo  {journal} {New Journal of Physics}\ }\textbf
  {\bibinfo {volume} {15}},\ \bibinfo {pages} {113016} (\bibinfo {year}
  {2013})}\BibitemShut {NoStop}%
\bibitem [{\citenamefont {Serafini}(2017)}]{serafini17QCV}%
  \BibitemOpen
  \bibfield  {author} {\bibinfo {author} {\bibfnamefont {A.}~\bibnamefont
  {Serafini}},\ }\href@noop {} {\emph {\bibinfo {title} {Quantum continuous
  variables: a primer of theoretical methods}}}\ (\bibinfo  {publisher} {CRC
  press},\ \bibinfo {year} {2017})\BibitemShut {NoStop}%
\bibitem [{\citenamefont {Peres}(1996)}]{peres96}%
  \BibitemOpen
  \bibfield  {author} {\bibinfo {author} {\bibfnamefont {A.}~\bibnamefont
  {Peres}},\ }\bibfield  {title} {\bibinfo {title} {Separability criterion for
  density matrices},\ }\href {https://doi.org/10.1103/PhysRevLett.77.1413}
  {\bibfield  {journal} {\bibinfo  {journal} {Physical Review Letters}\
  }\textbf {\bibinfo {volume} {77}},\ \bibinfo {pages} {1413} (\bibinfo {year}
  {1996})}\BibitemShut {NoStop}%
\bibitem [{\citenamefont {Weedbrook}\ \emph {et~al.}(2012)\citenamefont
  {Weedbrook}, \citenamefont {Pirandola}, \citenamefont
  {Garc{\'\i}a-Patr{\'o}n}, \citenamefont {Cerf}, \citenamefont {Ralph},
  \citenamefont {Shapiro},\ and\ \citenamefont {Lloyd}}]{weedbrook2012}%
  \BibitemOpen
  \bibfield  {author} {\bibinfo {author} {\bibfnamefont {C.}~\bibnamefont
  {Weedbrook}}, \bibinfo {author} {\bibfnamefont {S.}~\bibnamefont
  {Pirandola}}, \bibinfo {author} {\bibfnamefont {R.}~\bibnamefont
  {Garc{\'\i}a-Patr{\'o}n}}, \bibinfo {author} {\bibfnamefont {N.~J.}\
  \bibnamefont {Cerf}}, \bibinfo {author} {\bibfnamefont {T.~C.}\ \bibnamefont
  {Ralph}}, \bibinfo {author} {\bibfnamefont {J.~H.}\ \bibnamefont {Shapiro}},\
  and\ \bibinfo {author} {\bibfnamefont {S.}~\bibnamefont {Lloyd}},\ }\bibfield
   {title} {\bibinfo {title} {Gaussian quantum information},\ }\href
  {https://doi.org/10.1103/RevModPhys.84.621} {\bibfield  {journal} {\bibinfo
  {journal} {Reviews of Modern Physics}\ }\textbf {\bibinfo {volume} {84}},\
  \bibinfo {pages} {621} (\bibinfo {year} {2012})}\BibitemShut {NoStop}%
\end{thebibliography}%

%%%%%%%%%%%%%%%%%%%%%%%%%%%%%%%%%%%%%%%%%%%%%%%%%%%%%
\appendix

\begin{widetext}
%% SUPPLEMENTAL MATERIAL %%
\section*{Supplemental material} This supplemental material contains intermediate calculations omitted in the main text. Although this information can be derived  using the material provided in the main text, it may help readers unfamiliar with Gaussian quantum information to follow and reproduce our results---see Ref.~\cite{serafini17QCV} for an excellent account of Gaussian quantum information. In section \ref{sec:2}, we provide the main steps needed to compute the $S$-matrix corresponding to the white-black hole circuit of Fig.~3 of the main text. In section \ref{sec:3}, we show how to use this $S$-matrix to evolve a Gaussian state, and more concretely a one-mode squeezed state with thermal noise added to it. From this, we will derive some of the expressions used in the main body of the paper. In section \ref{sec:4} we provide some information about the effects of the environmental temperature on the entanglement produced during the Hawking process.

\subsection{White-Black hole S-matrix \label{sec:2}}
The $S$-matrix associated with the optical circuit of Fig.~3 of the main text can be constructed by combining two squeezers and two beam-splitters in the way indicated in Fig.~3. We will build the $S$-matrix first for creation and annihilation operators, and from it we will obtain the $S$-matrix for the quadrature operators by a simple change of variables. 

Let $\vec A_{in}=(a^{in}_{k_1},a^{in \dagger}_{k_1},a^{in}_{k_2},a^{in \dagger}_{k_2},a^{in}_{k_3},a^{in \dagger}_{k_3},a^{in}_{k_4},a^{in \dagger}_{k_4})$ be the (column) vector made of annihilation and creation operators for the $in$-modes, and let $\vec A_{out}$ be the corresponding vector for the $out$-modes. Time evolution   in Heisenberg's picture is encoded in the relation between  the $in$ and $out$ operators,  which can be written in matrix notation as:
\begin{equation}
\vec A_{out}= U^{\dagger} \vec A_{in}U=S_A \cdot  \vec{A}_{in}\, ,
\end{equation} 
where we have denoted with $S_A$ the $S$-matrix for annihilation and creation variables. 

In order to write $S_A$ from the circuit of Fig.~3 of the main text, we need to make a choice for the order of the modes in the intermediate steps of the circuit. We choose the order written in Fig.~\ref{order}. From this, and by recalling the action of a squeezer and a beam-splitter, written in Eqn.~(1) of the main text, it is straightforward to write the $S$-matrix corresponding to each element of the circuit of Fig.~3. Staring from left to right:

\begin{figure}[t]
{\centering     
\includegraphics[width = 1\textwidth]{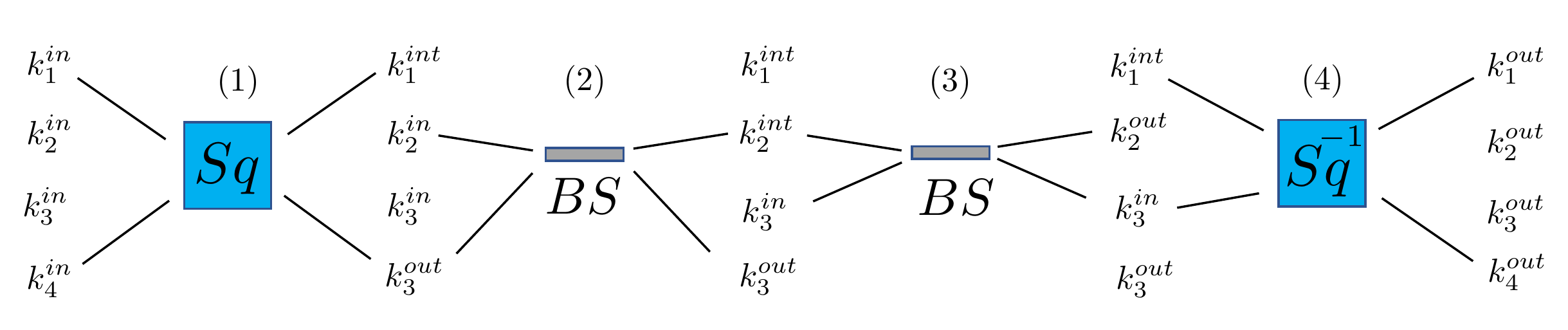}
}
  \caption{Schematic decomposition of the optical circuit of Fig.~3 of the main text, and our choice for the order of the modes in intermediate steps. The solid lines indicate the modes that are affected by each element of the circuit. The total $S$-matrix is the same regardless of the choice of the order in intermediate steps.} 
\label{order}
\end{figure}

\bea
S^{(1)}_{Sq} &=& \begin{pmatrix}
\cosh{r_H}&0&0&0&0&0&0&e^{i\phi}\sinh{r_H}\\ 0&\cosh{r_H}&0&0&0&0&e^{-i\phi}\sinh{r_H}&0\\
0&0&1&0&0&0&0&0\\0&0&0&1&0&0&0&0\\
0&0&0&0&1&0&0&0\\0&0&0&0&0&1&0&0\\ 0&e^{i\phi}\sinh{r_H}&0&0&0&0&\cosh{r_H}&0\\e^{-i\phi}\sinh{r_H}&0&0&0&0&0&0&\cosh{r_H}
\end{pmatrix} \nonumber \\ &=& \begin{pmatrix}
\cosh{r_H}\, \mathbb{I}_2& 0 & 0 & \sinh{r_H}\left(R_\phi\sigma_xR_\phi^\dagger\right) \\
0 & \mathbb{I}_2 & 0 & 0 \\
0 & 0 & \mathbb{I}_2 & 0 \\
\sinh{r_H}\left(R_\phi\sigma_xR_\phi^\dagger\right) & 0 & 0 & \cosh{r_H}\, \mathbb{I}_2
\end{pmatrix}
\eea
where in the second equality we have written $S^{(1)}_{Sq} $ in a more compact way, where $R_\phi=\e^{\im\, \sigma_z\, \phi/2}$,  $\sigma_x$ and $\sigma_z$ are Pauli matrices, and $\mathbb{I}_2$ is the two-dimensional identity.

\begin{equation}
S^{(2)}_{BS}=\begin{pmatrix} \mathbb{I}_2 & 0&0&0\\0&\cos{\theta}\,  \mathbb{I}_2&0&-\sin\theta\,  \mathbb{I}_2\\ 0&0&\mathbb{I}_2&0\\0&\sin\theta\,  \mathbb{I}_2&0&\cos{\theta}\,  \mathbb{I}_2\end{pmatrix}
\end{equation}

\begin{equation}
S^{(3)}_{BS}= \begin{pmatrix} \mathbb{I}_2 & 0&0&0\\0&\cos{\theta}\,  \mathbb{I}_2&\sin\theta\,  \mathbb{I}_2&0\\0&-\sin\theta\,  \mathbb{I}_2&\cos{\theta}\,  \mathbb{I}_2&0\\ 0&0&0&\mathbb{I}_2\end{pmatrix}
\end{equation}

\begin{equation}
 S^{(4)}_{Sq}=\begin{pmatrix}
\cosh{r_H}\, \mathbb{I}_2& 0  & -\sinh{r_H}\left(R_\phi\sigma_xR_\phi^\dagger\right)& 0 \\
0 & \mathbb{I}_2 & 0 & 0 \\
0 & 0  & 0& \mathbb{I}_2 \\
\sinh{r_H}\left(R_\phi\sigma_xR_\phi^\dagger\right) & 0 & \cosh{r_H}\, \mathbb{I}_2& 0 
\end{pmatrix}\end{equation}
The relative sign between some of the  elements of  the matrices $S^{(1)}_{Sq}$ and $S^{(4)}_{Sq}$ is due to the fact that the white hole squeezer is effectively the inverse of the black hole squeezer. The same argument explains the relative sign between elements of $S^{(2)}_{BS}$ and $S^{(3)}_{BS}$.

From these matrices we obtain $S_A$ as
\begin{equation}
    S_A= S^{(4)}_{Sq}\cdot S^{(3)}_{BS}\cdot S^{(2)}_{BS}\cdot S^{(1)}_{Sq}.
\end{equation}
It is straightforward to check that $ S_A$ is indeed a canonical transformation---as it should be---by checking that it leaves invariant the inverse of the symplectic matrix $\Omega^{-1}$: $S_A\cdot \Omega^{-1}\cdot S_A^{\top}=\Omega^{-1}$, where $\Omega^{-1}=\begin{pmatrix} 0&1\\-1&0\end{pmatrix} \otimes \mathbb{I}_{4}$. This guarantees that  time evolution preserves the canonical commutation relations. 

The $S$-matrix for the quadrature operators associated to each mode $x_i\equiv \frac{1}{\sqrt{2}} \, (a_{k_i}+a_{k_i}^{\dagger})$ and $p_i\equiv \frac{-i}{\sqrt{2}} \, (a_{k_i}-a_{k_i}^{\dagger})$ is obtained from  $S_A$ by 

\begin{equation}   S=M\cdot S_A\cdot M^{-1}
\end{equation}
where $M$ is the change-of-basis matrix 

\begin{equation}
M= \mathbb{I}_4\otimes\frac{1}{\sqrt{2}}\begin{pmatrix}
1&1\\
-i & i
\end{pmatrix}
\end{equation}
The matrix $S$ implements the Heisenberg evolution of the quadrature operators 
\begin{equation} \vec r_{in} \to \vec r_{out} =S\cdot  \vec r_{in}\end{equation} 
where $\vec r_{in}=(x_1^{in},p_1^{in},x_2^{in},p_2^{in},x_3^{in},p_3^{in},x_4^{in},p_4^{in})$, and $\vec r_{out}=(x_1^{out},p_1^{out},x_2^{out},p_2^{out},x_3^{out},p_3^{out},x_4^{out},p_4^{out})$.

\subsection{Evolution of Gaussian states \label{sec:3}}
Gaussian states are completely characterized by their first and second moments, $\vec \mu=\langle  \Psi |\vec r |\Psi\rangle$ and $\sigma=\langle \Psi| \{ (\vec r-\vec \mu),  (\vec r-\vec \mu)\}| \Psi \rangle$, where   $\{\cdot,\cdot\}$ is the symmetric anti-commutator, respectively ($\vec \mu$ is an 8-dimensional vector; $\sigma$ is an $8\times 8$ matrix).  All physical predictions can be written in terms of  $\vec \mu$ and $\sigma$. Therefore, rather than working with the (infinite dimensional) density matrix, it suffices to restrict attention to $\vec \mu$ and $\sigma$ when working with Gaussian states. For linear systems---for which the Hamiltonian is a second order polynomial of the canonical variables---time evolution preserves Gaussianty. Therefore, to obtain the way observable quantites evolve in time, it is sufficient to compute the time evolution of the first and second moments $\vec \mu$ and $\sigma$. This time evolution is obtained by a simple multiplication with the $S$-matrix, as follows: if  $\vec{\mu}_{in}$ and $\sigma_{in}$ are the mean and covariance of an $n$-mode Gaussian state $\Psi_{in}$, the evolution of the state under quadratic interactions is a Gaussian state with mean and covariance 
\begin{align}\label{out}
{\vec{\mu}}_{out}&= S\cdot \vec{\mu}_{in}\, ,\\
    \sigma_{out}&= S\cdot \sigma_{in} \cdot  S^{\top}\, .
\end{align}
In the main body of the paper, we use a one-mode squeezed-thermal Gaussian state as initial state, where the squeezing is on the mode $k^{in}_3$. This state is obtained by squeezing a thermal state with temperature $T_{env}$, and its mean and covariance are
\begin{align}
{\vec{\mu}}_{in}&= \vec 0\, ,\\
    \sigma_{in}&= (2 \, \bar n_{env}+1)
    \begin{pmatrix}
    \mathbb{I}_2 & 0 & 0 & 0 \\
    0 & \mathbb{I}_2 & 0 & 0 \\
    0 & 0 & e^{2r_I\sigma_z} & 0 \\
    0 & 0 & 0 & \mathbb{I}_2
    \end{pmatrix}
\end{align}
where $\bar n_{env}=(e^{\omega/T_{env}}-1)^{-1}$ is the mean number of thermal photons with frequency $\omega$ in the environment, and $r_I$ is the initial squeezing intensity. Notice that if $r_I$ =0 and $T_{env}=0$, the $in$ state reduces to the vacuum.

As described before, the mean and covariance of the final state are obtained by multiplying with $S$ (see Eqn.~\ref{out}). As explained in the main text, the effect of losses/inefficiencies can be included by replacing $\vec{\mu}_{out}\to \sqrt{\eta}\,  \vec{\mu}_{out}$ and 
$\sigma_{out}\to \eta, \sigma_{out}+(1-\eta)\, \mathbb{I}$, where $0\leq \eta \leq 1$ is the attenuation factor. From these quantities, we can compute all physical properties of the final state. 
For instance, the mean number of quanta can be obtained by applying the right hand side of the formula 
\begin{equation}
   \langle \hat{n} \rangle=\frac{1}{4}\text{Tr}\{\sigma_{red}\}+\vec{\mu}_{red}^\top\cdot\vec{\mu}_{red} -\frac{1}{2}N.\label{eqch1:particle_number_cov}
\end{equation}
to any (sub)system of output modes with reduced covariance matrix $\sigma_{red}$ and a vector of first moments $\vec{\mu}_{red}$. Here, Tr$\{\cdot\}$ denotes the action of trace and $N$ is the number of modes within the (sub)system.

\subsection{Temperature Conditions for entanglement degradation}\label{sec:4}

Assuming an initial ($in$) state of thermal equilibrium with an environment at temperature $T_{env}$, we analyze the entanglement for various bipartitions of the $out$ state of the white-black hole. We find that, when the environmental temperature is high enough (compared to the Hawking temperature, $T_H$, of the white-black hole), the entanglement between various various bipartitions vanishes entirely. From these observations, we set out to find exact conditions between the environment temperature $T_{env}$ and the Hawking temperature $T_H$, which must be satisfied in order for entanglement to be generated in the Hawking process. 

The temperature conditions that we discover are in one-to-one correspondence with entanglement conditions based on the positivity of partial transpose (PPT) criteria \cite{peres96} as applied to Gaussian states \cite{weedbrook2012}. We discover such conditions by computing the symplectic eigenvalues $\Tilde{\nu}$ \footnote{It is common to call $\nu$ the symplectic eigenvalues of the original covariance matrix $\sigma$. We reserve the notation $\tilde{\nu}$ for the matrix $\tilde{\sigma}$ -- the covariance matrix corresponding to a partially transposed quantum state. } of the matrix $\tilde{\sigma}_{out}$. The matrix $\tilde{\sigma}_{out}$ is found from the covariance matrix $\sigma_{out}$ by first choosing a bipartition of modes, $A|B$, and then taking the $B$ (or $A$) modes and flipping the sign of the $p$ quadrature for that subsystem of modes. Note that, for Gaussian states, the condition for there to exist entanglement between subsystems $A$ and $B$ is that at least one symplectic eigenvalue of $\tilde{\sigma}_{out}$ satisfies $\Tilde{\nu}<1$. It is precisely this inequality that our temperature conditions in Table \ref{table:table1} correspond to. 

In what follows, we shall ignore the $k_2^{out}$ mode from our analysis since, to a very good approximation, it decouples entirely from all other modes. This is because the only elements in the white-black hole circuit which mixes $k_2^{out}$ with the other modes are the beam-splitters, and we find numerically that their transmission amplitudes are $\cos\theta\approx1$, as can be seen from Fig.~4 of the main body of this paper. 
We are thus left with a system of three modes $k_1^{out}$, $k_3^{out}$, and $k_4^{out}$, for which a non-zero value of the logarithmic negativity is a \textit{necessary and sufficient} condition for entanglement \cite{serafini17QCV}. 

The entanglement conditions for various sub-systems are generally frequency- dependent, and the final expressions are quite lengthy and not illuminating. For clarity, we quote the results in the low-particle ($\omega/T_H\rightarrow\infty$) and high-particle ($\omega/T_H\rightarrow0$) limits in Table \ref{table:table1}, for two different bipartitions: (i) the mode pair containing the most amount of entanglement out of all other mode pairs -- the correlated output modes of the white hole, $(k_1^{out},k_4^{out})$ and (ii) the bipartition containing the most amount of entanglement out of all other possible bipartitions, $(k_1^{out}|k_3^{out},k_4^{out})$. Observe that there is entanglement for the mode pair $(k_1^{out}|k_4^{out})$ when $T_{env}<2T_H$ at high $\omega/T_H$. This is similar to the condition found in reference \cite{bruschi2013}.
We see obvious deviations from this condition here, since, at low $\omega/T_H$, the condition becomes $T_{env}<T_H$. 
Furthermore, we see that the entanglement in the multi-mode bipartition $(k_1^{out}|k_3^{out},k_4^{out})$ is much more robust to ambient thermal noise than the entanglement in the mode pair $(k_1^{out}|k_4^{out})$, as can be seen by the corresponding temperature conditions in Table \ref{table:table1}.

\begin{table}[t]
		\renewcommand{\arraystretch}{1.6}
		\centering
		\begin{tabular}{c c c}
			\hline\hline Sub-systems &$(\omega/T_H\rightarrow \infty)$ & $(\omega/T_H\rightarrow 0)$\\ \hline
			$(k_1^{out}|k_4^{out})$  & $T_{env}<2T_H$ & $T_{env}<T_H$
			\\
			$(k_1^{out}|k_3^{out},k_4^{out}) \ \ $   & $T_{env}<\frac{2T_H}{1-\ln(2)T_H/\omega}$ &\hspace{.5em} $T_{env}<4T_H(\frac{T_H}{\omega})$
			\\
			\hline\hline
		\end{tabular}
\caption{\textbf{Entanglement conditions for two different bipartitions of the white-black hole initially at thermal equilibrium at temperature $T_{env}$ (zero initial squeezing).}}\label{table:table1}
\end{table}

\end{widetext}
\end{document}